\documentclass[journal]{IEEEtran}
\usepackage[latin9]{inputenc}
\usepackage{float}
\usepackage{amsmath}
\usepackage{amssymb}
\usepackage{graphicx}

\makeatletter

\providecommand{\tabularnewline}{\\}
\floatstyle{ruled}
\newfloat{algorithm}{tbp}{loa}
\providecommand{\algorithmname}{Algorithm}
\floatname{algorithm}{\protect\algorithmname}

\usepackage{hyperref}
\hypersetup{
    colorlinks=true,
    linkcolor=blue,
    filecolor=magenta,      
    urlcolor=cyan,
}

\ifCLASSINFOpdf
\else
\fi
\usepackage{algorithm}
\usepackage{algpseudocode}

\makeatother

\begin{document}

\title{Reliable Over-the-Air Computation by Amplify-and-Forward Based Relay }
\author{\IEEEauthorblockN{Suhua TANG$^{1}$, Huarui YIN$^{2}$, Chao ZHANG$^{3}$,
Sadao OBANA$^{1}$} \\
\IEEEauthorblockA{ $^{1}$Graduate School of Informatics and Engineering,
The University of Electro-Communications, Japan \\
 $^{2}$Dept. of Electronic and Information Science, University of
Science and Technology of China, China\\
$^{3}$School of Information and Communications Engineering, Xi\textquoteright an
Jiaotong University, China\\
 Email: shtang@uec.ac.jp}}
\maketitle
\begin{abstract}
In typical sensor networks, data collection and processing are separated.
A sink collects data from all nodes sequentially, which is very time
consuming. Over-the-air computation, as a new diagram of sensor networks,
integrates data collection and processing in one slot: all nodes transmit
their signals simultaneously in the analog wave and the processing
is done in the air. This method, although efficient, requires that
signals from all nodes arrive at the sink, aligned in signal magnitude
so as to enable an unbiased estimation. For nodes far away from the
sink with a low channel gain, misalignment in signal magnitude is
unavoidable. To solve this problem, in this paper, we investigate
the amplify-and-forward based relay, in which a relay node amplifies
signals from many nodes at the same time. We first discuss the general
relay model and a simple relay policy. Then, a coherent relay policy
is proposed to reduce relay transmission power. Directly minimizing
the computation error tends to over-increase node transmission power.
Therefore, the two relay policies are further refined with a new metric,
and the transmission power is reduced while the computation error
is kept low. In addition, the coherent relay policy helps to reduce
the relay transmission power by half, to below the limit, which makes
it one step ahead towards practical applications. 
\end{abstract}

\begin{IEEEkeywords}
Over-the-air computation, amplify and forward, relay
\end{IEEEkeywords}

\section{Introduction}


Many sensor nodes will be deployed to sense the environment so as
to support context-aware applications in the future smart society.
These sensors will be connected to the Internet via techniques such
as NB-IoT and LoRa \cite{Sinha17}. In the data collection process,
generally, the sink node has to collect data from each node, one by
one, which will take a long time when there are millions of nodes
in the coverage of a sink node. In addition, many nodes share a common
channel, and the increase in the number of nodes will lead to more
transmission collisions.

On the other hand, in some tasks, people are only interested in the
statistics of sensor data, e.g., the average temperature or moisture
in an area, instead of their respective values. For these cases, it
is possible to exploit a more efficient method called over-the-air
computation (AirComp) \cite{Nazer07}. Typically, all nodes simultaneously
transmit their signals in the analog wave \cite{Xiao08}, and the
data collection and processing are integrated in one slot. Then, their
fusion (sum) is computed by the superposition of electromagnetic waves
in the air, at the antenna of the sink. An essential feature of AirComp
is the uncoded analog transmission, which seems inferior to digital
transmissions. Actually, it is proved that the computation error in
AirComp based estimation is exponentially smaller than the digital
schemes when using the same amount of resources \cite{CZLee20}. Besides
the sum operation, AirComp can support any kind of nomographic functions
\cite{Buck79,Goldenbaum13c,Abari16}, if only proper pre-processing
is done at the sensor nodes and post-processing is done at the sink.
Recently, deep AirComp is studied, using deep neural networks in the
pre-processing and post-processing, which enables more advanced processing
of sensor data \cite{HaoYe20}.

To ensure an unbiased data fusion, it is required that signals from
all nodes arrive at the sink, aligned in signal magnitude. This is
usually achieved by transmission power control at sensor nodes \cite{Liu20,Cao19}.
Specifically, each node uses a transmission power inversely proportional
to the channel gain so as to mitigate the difference in channel gains.
Obviously, for nodes far away from the sink with a low channel gain,
even using the largest transmission power cannot equalize the channel,
and the misalignment in signal magnitude unavoidably occurs under
the constraint of transmission power, which can be regarded as an
outage. 

Path diversity by a relay is a conventional and effective method to
reducing the outage probability. The decode-and-forward (DF) method
applies codes to protect signals. Amplify-and-forward (AF) is simpler,
where a relay node simply amplifies the received signal (together
with noise). There have been many literature on relay for the unicast
communication, either AF \cite{Zhao06,Yang07}, DF \cite{Wang07},
or their comparison \cite{Souryal06}. In addition, network coding-based
relay also has been studied for the bidirectional communication \cite{Popovski07}
and the multiple access channel \cite{Tang09}. In the multiple access
channel, compute-and-forward \cite{Nazer11} works in a similar way
as network coding, where a relay node decodes the linear combination
of multiple received messages and forwards it towards the sink. The
sink, with efficient equations of messages, can solve each message
separately. But these relay methods cannot be directly applied to
AirComp. Recently, intelligent reflecting surface (IRS) is suggested
for AirComp \cite{Jiang19}, where a reflection surface, as a passive
relay node, is used to shape the phase of each signal. This may be
possible for the mmWave band (or higher frequencies) where electromagnetic
wave can propagate directionally along desired paths, but it is difficult
to apply IRS in a reflection-rich environment for the typical IoT
frequency bands (e.g., 920MHz, 2.4GHz).

In this paper, we will investigate how to use relay, more specifically,
AF-based relay, to improve the performance of AirComp. To the best
of our knowledge, this is the first work on AF-based AirComp. AF is
considered because signals in AirComp are transmitted in the analog
wave. In the communication, the relay node will amplify signals from
many nodes and forward them to the sink, and the whole process should
try to ensure the alignment of signal magnitudes at the sink so as
to reduce the computation error. 

The contribution of this paper is three-fold, as follows:
\begin{itemize}
\item We first present the general relay model for AirComp, and investigate
a simple relay (SimRelay) policy, in which a node either directly
transmits its signal to the sink or via the relay, but not both. Then,
we point out the problem: relay transmission power increases with
the number of nodes that use the relay node.
\item To reduce relay transmission power, we propose a coherent relay (CohRelay)
policy, in which a node can divide its power to transmit its signal
to both the relay and the sink, and the replicas of its signal are
coherently combined together at the sink. We also investigate the
impact of the number of nodes using the relay node.
\item We discuss the tradeoff between computation error and transmission
power. The computation error is composed of signal part  and noise
part. We find that directly minimizing the computation error may lead
to a large increase in node transmission power when the noise part
is dominant. Therefore, we further refine the relay policies, avoiding
over-reducing the noise part. This has little impact on the computation
error, but greatly reduces node transmission power. 
\end{itemize}
Extensive simulation evaluations confirm the effectiveness of the
proposed methods. Especially, CohRelay greatly reduces the relay transmission
power to below the limit, which makes it one step ahead towards practical
applications. 

In the rest of this paper, in Sec.\ref{sec:Related}, we review the
AirComp model and previous work on improving its performance. Then,
in Sec.\ref{sec:Relay-Model}, we present the relay model for AirComp,
and investigate two relay policies. With some simulation results,
we illustrate the problem of over increase in node transmission power.
Then, the two relay policies are refined and evaluated in Sec.\ref{sec:Refine}.
Finally, in Sec.\ref{sec:Conclusion}, we conclude this paper and
point out future work.

\section{Related Work\label{sec:Related}}

Here, we review the AirComp method and previous solutions to channel
fading. 

\subsection{Basic over-the-air computation}

We first introduce the basic AirComp model \cite{Liu20}.  A sensor
network is composed of $K$ sensor nodes and 1 sink. The sensing result
at the $k_{th}$ node is represented by the signal $x_{k}\in[-v,v]\in\mathbb{C}$,
which has zero mean and unit variance ($E(|x_{k}^{2}|)=1$). The sink
will compute the sum of sensing data from all nodes. Both the nodes
and the sink have a single antenna. To overcome channel fading, the
$k_{th}$ node pre-amplifies its signal by a Tx-scaling factor $b_{k}\in\mathbb{C}$.
The channel coefficient between sensor $k$ and the sink is $h_{k}\in\mathbb{C}$.
The sink further applies a Rx-scaling factor $a\in\mathbb{C}$ to
the received signal, as follows

\begin{equation}
s=a\cdot\left(\sum_{k=1}^{K}h_{k}b_{k}x_{k}+n\right),\label{eq:1slot-model}
\end{equation}
where $n\in\mathbb{C}$ is the additive white Gaussian noise (AWGN)
at the sink with zero mean and power being $\sigma^{2}$. It is assumed
that channel coefficient $h_{k}$ is known by both node $k$ and the
sink. Then, in a centralized way, the sink can always adjust $b_{k}$
to ensure that $h_{k}b_{k}$ is real and positive. Therefore, in the
following, it is assumed that $h_{k}\in\mathbb{R^{+}}$, $b_{k}\in\mathbb{R^{+}}$,
and $a\in\mathbb{\mathbb{R^{+}}}$ for the simplicity of analysis. 

The computation error is defined as the mean squared error (MSE) between
the received signal sum $s$ and the target signal $\sum_{k=1}^{K}x_{k}$,
as follows (by using the facts that signals are independent of each
other and independent of noise, $E(x_{k})=0$, $E(|x_{k}^{2}|)=1$) 

\begin{align}
MSE & \negmedspace=\negmedspace E\left\{ |s-\negmedspace\sum_{k=1}^{K}x_{k}|^{2}\right\} =\negmedspace\sum_{k=1}^{K}|ah_{k}b_{k}-1|^{2}+\sigma^{2}|a|^{2}.\label{eq:mse-orig}
\end{align}

With the maximal power constraint, $|b_{k}x_{k}|^{2}$ should be no
more than $P'$, the maximal power. Let $P_{max}$ denote $P'/v^{2}$.
Then, we have $b_{k}^{2}\le P'/v^{2}=P_{max}$.  By sorting the channel
coefficient ($h_{k}$) in the increasing order, the optimal solution
(under the power constraint $|b_{k}|^{2}\le P_{max},k=1,\cdots,K$)
depends on a critical number, $i^{\star}$ \cite{Liu20}. A node whose
index is below $i^{\star}$ uses the maximal power $P_{max}$, and
otherwise uses a power inversely proportional to the channel gain.
Then, the computation MSE is computed as follows:

\begin{align}
MSE & =\sum_{k=1}^{i^{\star}}|ah_{k}\sqrt{P_{max}}-1|^{2}+\sigma^{2}|a|^{2}.\label{eq:mse-twc20}
\end{align}

This computation MSE may be caused by channel fading or noise. The
former decides the error in the signal magnitude of $i^{\star}$ weak
signals and the latter decides the term $\sigma^{2}|a|^{2}$. 

\subsection{Previous improvement on AirComp}

When some nodes are far away from the sink, the magnitudes of their
signals cannot be aligned with that of other signals from nearer nodes.
Some efforts have been devoted to solving this problem. The work in
\cite{Liu20} studies the power control policy, aiming to minimize
the computation error by jointly optimizing the transmission power
and a receive scaling factor at the sink node. Generally, the principle
of channel inversion  is adopted. Specifically, with the common signal
magnitude being $\alpha$ ($\alpha=1/a$), the transmission power
of node $k$ is computed as $b_{k}=min\{\alpha/h_{k},\sqrt{P_{max}}\}$,
being the former if $\alpha/h_{k}$ is below the power constraint,
and otherwise, using the maximal power. In \cite{Cao19}, the authors
further consider the time-varying channel by regularized channel inversion,
aiming at a better tradeoff between the signal-magnitude alignment
and noise suppression. Antenna array was also investigated in \cite{Zhu19H,Wen19}
to support vector-valued AirComp. 

AirComp is an efficient solution in federated learning, where the
model update is to be transmitted from each node to the common sink,
aggregated there, and then sent back to each node for future data
processing. Specific consideration on AirComp is also studied. Because
information from some of the nodes is sufficient, node selection based
on the channel gain is suggested in \cite{Amiri20}, although this
does not apply to general AirComp where signals from all nodes are
needed. Sery \emph{et al.} further suggests precoding and scaling
operations to gradually mitigate the effect of the noisy channel so
as to facilitate the convergence of the learning process \cite{TomerSery20}.

\section{AirComp with AF-based relay \label{sec:Relay-Model}}

A wireless signal attenuates as the propagation distance increases.
With a single antenna, the effect of transmission power control in
dealing with path loss and channel fading is limited. Therefore, we
try to exploit relay, which has been proven to be effective in conventional
unicast communications.

\subsection{System framework\label{sec:Framework}}

\begin{figure}
\centering

\includegraphics[width=6cm]{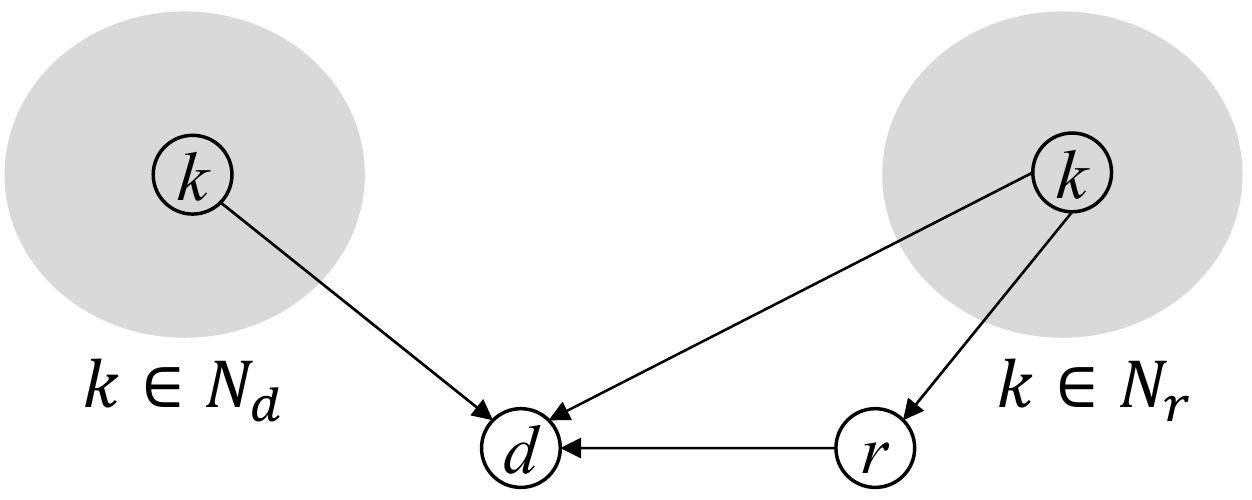}

\caption{\label{fig:RelayModel}Relay for AirComp, an analogy to a conventional
relay network.}
\end{figure}

The network consists of $K$ sensor nodes, a relay $r$ and a sink
$d$. Sink $d$ will compute the sum of sensing data from all nodes,
via the help of relay $r$. All nodes, relay $r$ and sink $d$ use
a single antenna. Each node has a constraint of the maximal transmission
power. But we assume that the relay has no constraint of transmission
power, and investigate how much power is required for relaying signals.
 Nodes near to the sink can directly communicate with the sink, while
nodes farther away can rely on the relay to help. Then, all nodes
are divided into two groups. A node $k$ is either  a neighbor of
$r$ ($k\in N_{r}$) and will use relay $r$, or a non-neighbor of
$r$ ($k\in N_{d}$) and will directly transmit its signal to sink
$d$. Fig.\ref{fig:RelayModel} shows an analogy to a conventional
relay network. The difference is that there are more than one node
in $N_{r}$ and $N_{d}$.

\begin{figure}
\centering

\includegraphics[width=9cm]{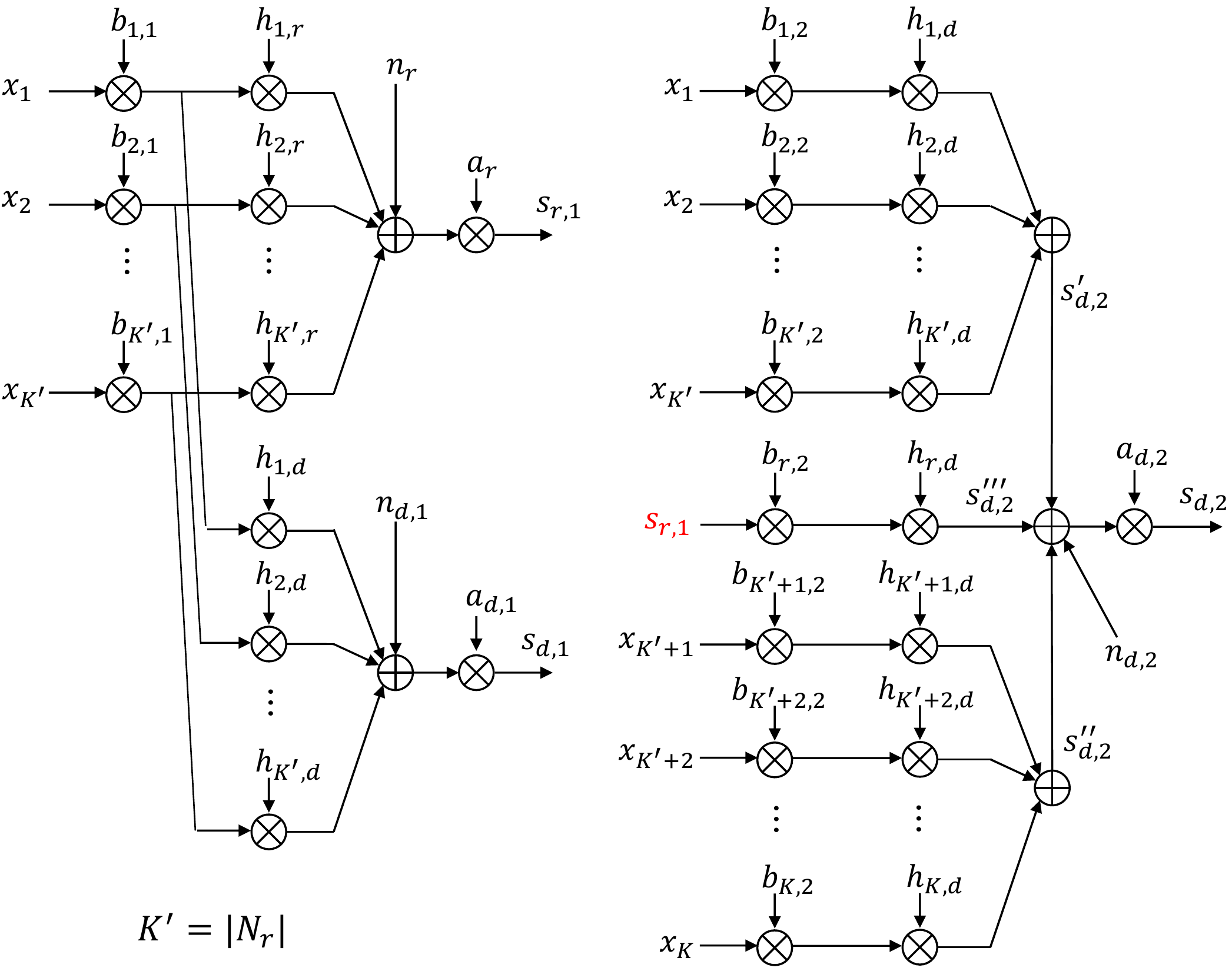}

\caption{\label{fig:RelayModelDetail}Detailed relay model for AirComp. $K$
sensor nodes each have a signal to send to the common sink. Nodes
in $N_{r}$ transmit their signals to the sink via the relay, while
nodes in $N_{d}$ directly transmit their signals to the sink.}
\end{figure}

We assume that (i) the sensor network is fixed without node mobility,
and channel coefficients ($h_{k,r}\in\mathbb{C}$ and $h_{k,d}\in\mathbb{C}$,
representing channel coefficients from node $k$ to relay $r$ and
sink $d$, respectively) are constant within a period of time, (ii)
each node (and relay $r$) knows channel coefficients of its links
to sink $d$ and relay $r$, and (iii) channel coefficients of all
links are known to sink $d$\footnote{It is common to assume that the sink knows the CSI of node-relay links
and node-sink links in the AF-based relay control \cite{Zhao06} and
the IRS-based relay control \cite{Jiang19}. }, which decides the node grouping policy (decides which nodes to use
the relay) and other parameters.

Similar to the conventional AF method, the whole transmission is divided
into two slots. But the transmission powers (Tx-scaling factor $b_{k,1}\in\mathbb{C}$
and $b_{k,2}\in\mathbb{C}$ in two time slots) are adjusted per node
per slot. The detailed process is shown in Fig.\ref{fig:RelayModelDetail}.

In the first slot, a neighbor node ($k\in N_{r}$) of relay $r$ transmits
its signal using a Tx-scaling factor $b_{k,1}$. The signals received
at relay $r$ and sink $d$ are 

\begin{equation}
s_{r,1}=a_{r,1}\cdot\left(\sum_{k\in N_{r}}h_{k,r}b_{k,1}x_{k}+n_{r,1}\right),
\end{equation}

\begin{equation}
s_{d,1}=a_{d,1}\cdot\left(\sum_{k\in N_{r}}h_{k,d}b_{k,1}x_{k}+n_{d,1}\right),
\end{equation}
where $a_{r,1}\in\mathbb{C}$ and $a_{d,1}\in\mathbb{C}$ are Rx-scaling
factors, and $n_{r,1}$ and $n_{d,1}$ are AWGN noises with zero mean
and variance being $\sigma^{2}$.

In the second slot, all nodes transmit their signals to sink $d$,
and node $k$ uses a Tx-scaling factor $b_{k,2}$. Meanwhile relay
$r$ also forwards its received signal, using a Tx-scaling factor
$b_{r,2}$. Signals arriving at sink $d$ are composed of 3 parts,
as follows:

\begin{equation}
s'_{d,2}=\sum_{k\in N_{r}}h_{k,d}b_{k,2}x_{k},
\end{equation}

\begin{equation}
s''_{d,2}=\sum_{k\in N_{d}}h_{k,d}b_{k,2}x_{k},
\end{equation}

\begin{equation}
s'''_{d,2}=h_{r,d}b_{r,2}s_{r,1},
\end{equation}
where $s'_{d,2}$ is the signal from $k\in N_{r}$, $s''_{d,2}$ is
the signal from $k\in N_{d}$, and $s'''_{d,2}$ is the relayed signal.
Then, the overall signal at the second slot is

\begin{equation}
s_{d,2}=a_{d,2}\cdot\left(s'_{d,2}+s''_{d,2}+s'''_{d,2}+n_{d,2}\right),
\end{equation}
where $a_{d,2}\in\mathbb{C}$ is a Rx-scaling factor, and $n_{d,2}$
is AWGN noise with zero mean and variance being $\sigma^{2}$.

Sink $d$ adds the signals received in the two slots. For a signal
from a neighbor ($k\in N_{r}$) of relay $r$, its overall coefficient
at the sink is

\begin{align}
\beta_{k} & =a_{d,1}h_{k,d}b_{k,1}+a_{d,2}h_{k,d}b_{k,2}\nonumber \\
 & +a_{d,2}h_{r,d}b_{r,2}\cdot a_{r,1}h_{k,r}b_{k,1},k\in N_{r}.
\end{align}
Its first term corresponds to the signal directly received in the
first slot, its second term corresponds to the signal directly received
in the second slot, and its third term corresponds to the relayed
signal. 

For a signal from a node not a neighbor ( $k\in N_{d}$) of relay
$r$, its coefficient at sink $d$ is

\begin{equation}
\beta_{k}=a_{d,2}h_{k,d}b_{k,2},k\in N_{d}.
\end{equation}
The overall noise is 

\begin{equation}
n_{a}=a_{d,1}n_{d,1}+a_{d,2}n_{d,2}+a_{d,2}h_{r,d}b_{r,2}\cdot a_{r,1}n_{r,1}.
\end{equation}
All the parameters are to be solved by minimizing the following computation
MSE (under the power constraint $|b_{k,1}|^{2}\le P_{max},|b_{k,2}|^{2}\le P_{max},k=1,\cdots,K$).

\begin{align}
MSE & \negmedspace=\negmedspace E\left\{ |\sum_{k=1}^{K}\beta_{k}x_{k}\negmedspace+\negmedspace n_{a}\negmedspace-\negmedspace\sum_{k=1}^{K}x_{k}|^{2}\right\} \negmedspace=\negmedspace\sum_{k=1}^{K}|\beta_{k}-1|^{2}\nonumber \\
 & +\sigma^{2}(|a_{d,1}|^{2}+|a_{d,2}|^{2}+|a_{d,2}h_{r,d}b_{r,2}a_{r,1}|^{2}).\label{eq:mse-relay}
\end{align}
It is difficult to directly solve this problem. In the following,
we discuss its solution under several relay policies.

\subsection{Nodes grouping and relay transmission power}

Different from a conventional relay method, relay $r$ has to amplify
$|N_{r}|$ signals, and the overall signal to be relayed is 

\begin{equation}
\sum_{k\in N_{r}}a_{r,1}h_{k,r}b_{k,1}x_{k}.
\end{equation}
To ensure that all signals are aligned in magnitude at sink $d$,
the magnitude of the relayed signals ($a_{d,2}h_{r,d}b_{r,2}\cdot a_{r,1}h_{k,r}b_{k,1}$,
$k\in N_{r}$) should approach that of directly received signals ($a_{d,2}h_{k,d}b_{k,2}$,
$k\in N_{d}$). Here, relay $r$ will use a Tx-scaling factor $b_{r,2}$
which depends on node transmission power ($b_{k,1}$), channel gains
($h_{r,d}$, $h_{k,r}$), and alignment with other nodes ($a_{d,2}$,
$a_{r,1}$). The transmission power required at the relay node is

\begin{equation}
|b_{r,2}|^{2}\cdot\sum_{k\in N_{r}}|a_{r,1}h_{k,r}b_{k,1}|^{2}.\label{eq:relay-pow}
\end{equation}
Obviously, the relay transmission power linearly increases with the
number of nodes using the relay, which is a big problem. Therefore,
it is impractical to use relay for all nodes.

To solve this problem, we propose that nodes far away from sink $d$
while near to relay $r$ should use the relay. Therefore, nodes are
sorted in the ascending order of $|h_{k,d}|^{2}-|h_{k,r}|^{2}$, the
difference of channel gain to sink $d$ and relay $r$. The top nodes
will use relay $r$, and the percentage of nodes using relay $r$
is a parameter. 

\subsection{Simple Relay Policy}

We first consider a simple relay (SimRelay) policy. $s_{d,1}$ is
neglected ($a_{d,1}=0$) and $s'_{d,2}$ is not transmitted ($b_{k,2}=0,k\in N_{r}$).
In other words, in the first slot, signals from $k\in N_{r}$ are
sent to relay $r$, and in the second slot, signals from $k\in N_{d}$
are directly sent to sink $d$ and signals from $k\in N_{r}$ are
forwarded to sink $d$ by relay $r$. This is the most simple relay
method: the direct link is neglected once the relay is used. 

With $a_{d,1}=0$, $b_{k,2}=0$ ($k\in N_{r}$), and $a_{d,2}h_{r,d}b_{r,2}=c$,
the computation MSE in Eq.(\ref{eq:mse-relay}) can be rewritten as 

\begin{align}
MSE & =\sum_{k\in N_{r}}|(ca_{r,1})h_{k,r}b_{k,1}-1|^{2}+\sigma^{2}|ca_{r,1}|^{2}\nonumber \\
 & +\sum_{k\in N_{d}}|a_{d,2}h_{k,d}b_{k,2}-1|^{2}+\sigma^{2}|a_{d,2}|^{2}.
\end{align}
Because $c$ can be merged into $a_{r,1}$, we denote their product
as $a'_{r,1}=ca_{r,1}$, and the computation MSE can be computed as
the sum of 

\begin{align}
MSE_{r} & =\sum_{k\in N_{r}}|a'_{r,1}h_{k,r}b_{k,1}-1|^{2}+\sigma^{2}|a'_{r,1}|^{2},\nonumber \\
MSE_{d} & =\sum_{k\in N_{d}}|a_{d,2}h_{k,d}b_{k,2}-1|^{2}+\sigma^{2}|a_{d,2}|^{2}.\label{eq:mse-2aircomp}
\end{align}
Because $MSE_{r}$ and $MSE_{d}$ depend on different nodes and different
parameters, the relay problem is equivalent to two AirComp problems,
one from $k\in N_{r}$ to relay $r$ in the first slot, and the other
from $k\in N_{d}$ to sink $d$ in the second slot. Each can be solved
by using the power control algorithm suggested in \cite{Liu20}. Because
$b_{k,1}$ and $b_{k,2}$ can be adjusted to ensure $h_{k,r}b_{k,1}$
and $h_{k,d}b_{k,2}$ are positive real numbers, in the analysis,
$h_{k,r}\in\mathbb{R^{+}}$, $h_{k,d}\in\mathbb{R^{+}}$, $b_{k,1}\in\mathbb{R^{+}}$,
$b_{k,2}\in\mathbb{R^{+}}$, $a'_{r,1}\in\mathbb{R^{+}}$, $a_{d,2}\in\mathbb{R^{+}}$
are assumed for the simplicity of analysis.

In SimRelay, relay $r$ has to amplify the whole signals from $|N_{r}|$
nodes, which requires much transmission power.

\subsection{Coherent Relay Policy}

To reduce relay transmission power, we consider a coherent relay (CohRelay)
policy. A node using relay divides its power into two parts, and transmits
its signal twice. Sink $d$ receives two coherent copies of the same
signal and adds them together.

In this case, $s_{d,1}$ is neglected ($a_{d,1}=0$) but $s'_{d,2}$
($k\in N_{r}$) is transmitted. Compared with SimRelay, the difference
is that in the second slot, nodes $k\in N_{r}$ transmit their signals
again. With $a_{d,1}=0$, $a_{d,2}h_{r,d}b_{r,2}=c$, and $a'_{r,1}=ca_{r,1}$,
the computation MSE in Eq.(\ref{eq:mse-relay}) can be rewritten as 

\begin{align}
MSE & =\sum_{k\in N_{r}}|a_{d,2}h_{k,d}b_{k,2}+a'_{r,1}h_{k,r}b_{k,1}-1|^{2}+\sigma^{2}|a'_{r,1}|^{2}\nonumber \\
 & +\sum_{k\in N_{d}}|a_{d,2}h_{k,d}b_{k,2}-1|^{2}+\sigma^{2}|a_{d,2}|^{2}.\label{eq:mse-relay2}
\end{align}
Because $a_{d,2}$ also appears in the first sum, this cannot be simply
divided into two AirComp problems like SimRelay. But $h_{k,r}\in\mathbb{R^{+}}$,
$h_{k,d}\in\mathbb{R^{+}}$, $b_{k,1}\in\mathbb{R^{+}}$, $b_{k,2}\in\mathbb{R^{+}}$,
$a'_{r,1}\in\mathbb{R^{+}}$, $a_{d,2}\in\mathbb{R^{+}}$ can be assumed
in the analysis. Then, $a_{d,2}h_{k,d}b_{k,2}$ in the first sum is
a positive real number. Without this term, like SimRelay, an initial
estimation of $a'_{r,1}$ and $a_{d,2}$ can be computed, by minimizing
$MSE_{r}$ and $MSE_{d}$ in Eq.(\ref{eq:mse-2aircomp}), respectively. 

Next consider the presence of $a_{d,2}h_{k,d}b_{k,2}$ in the first
sum of Eq.(\ref{eq:mse-relay2}). Assume originally some $a'_{r,1}$
and $b_{k,1}$ make $a'_{r,1}h_{k,r}b_{k,1}$ equal to 1.0 (or approach
1 under the maximal power constraint). If $b_{k,1}$ is fixed, the
presence of $a_{d,2}h_{k,d}b_{k,2}$ (a positive number) makes it
possible to use a smaller $a'_{r,1}$ to make $a_{d,2}h_{k,d}b_{k,2}+a'_{r,1}h_{k,r}b_{k,1}$
reach 1.0. Meanwhile, the term $\sigma^{2}|a'_{r,1}|^{2}$ also decreases.
In other words, it is possible to decrease $a'_{r,1}$ in a certain
range to reduce the first sum in Eq.(\ref{eq:mse-relay2}). Therefore,
a heuristic algorithm is to use the initial estimation of $a'_{r,1}$
as a seed, and then gradually decrease it to find the minimum while
fixing $a_{d,2}$ (ensuring the minimum of the second sum in Eq.(\ref{eq:mse-relay2})).

Actually, $b_{k,1}$ and $b_{k,2}$ depend on the setting of $a'_{r,1}$
and $a_{d,2}$. In addition, to ensure a fair comparison with SimRelay,
it is assumed that the overall power, $|b_{k,1}|^{2}+|b_{k,2}|^{2}=P$,
should be no more than $P_{max}$. Then, the power allocation for
$b_{k,1}$ and $b_{k,2}$ ($k\in N_{r}$) is to maximize the term
$a_{d,2}h_{k,d}b_{k,2}+a'_{r,1}h_{k,r}b_{k,1}$, under the power constraint.
According to the Cauchy--Schwarz inequality \cite{Steele04} 

\begin{align}
 & \left((a'_{r,1}h_{k,r})b_{k,1}+(a_{d,2}h_{k,d})b_{k,2}\right)^{2}\nonumber \\
 & \le\left((a'_{r,1}h_{k,r})^{2}+(a_{d,2}h_{k,d})^{2}\right)\left(b_{k,1}^{2}+b_{k,2}^{2}\right),
\end{align}
and the equality holds if and only if 

\begin{equation}
\frac{b_{k,1}}{a'_{r,1}h_{k,r}}=\frac{b_{k,2}}{a_{d,2}h_{k,d}}=\rho_{k}.\label{eq:case2-tpc}
\end{equation}
Then, with $|b_{k,1}|^{2}+|b_{k,2}|^{2}=P\le P_{max}$, $\rho_{k}$
can be computed as 

\begin{equation}
\rho_{k}(P)=\frac{\sqrt{P}}{\sqrt{\left(a'_{r,1}h_{k,r}\right)^{2}+\left(a_{d,2}h_{k,d}\right)^{2}}}.
\end{equation}
On this basis, $b_{k,1}$ and $b_{k,2}$ are computed from Eq.(\ref{eq:case2-tpc}),
and the value of $a_{d,2}h_{k,d}b_{k,2}+a'_{r,1}h_{k,r}b_{k,1}$ is
computed as

\begin{align}
\gamma_{k}(P) & =\rho_{k}(P)\cdot\left((a'_{r,1}h_{k,r})^{2}+(a_{d,2}h_{k,d})^{2}\right).
\end{align}
If $\gamma_{k}(P_{max})$ is greater than 1.0, setting $\gamma_{k}(P)=1$
can find $\rho_{k}(P)$ and the powers ($b_{k,1}$ and $b_{k,2}$)
that lead to 0 error in the signal magnitude. 

The whole process of finding optimal parameters and the corresponding
computation MSE is described in Algorithm \ref{alg:relay2}.

\begin{algorithm}
\begin{algorithmic}[1]

\Procedure{FindParamForCohRelay}{$h_{k,r}$, $h_{k,d}$}

\State Initialize $a_{d,2}$, by minimizing $MSE_{d}$ in Eq.(\ref{eq:mse-2aircomp}) 

\State Initialize $a'_{r,1}$, by minimizing $MSE_{r}$ in Eq.(\ref{eq:mse-2aircomp}) 

\State $MSE_{rd}$ = OneIter($h_{k,r}$, $h_{k,d}$, $a'_{r,1}$,
$a_{d,2}$)  

\While {$True$} \Comment{Iteration}

\State $MSE'_{rd}$ = OneIter($h_{k,r}$, $h_{k,d}$, $a'_{r,1}\negmedspace-\negmedspace\delta$,
$a_{d,2}$)

\If {$MSE'_{rd}<MSE_{rd}$}

\State $a'_{r,1}\leftarrow a'_{r,1}-\delta$ \Comment{Update $a'_{r,1}$}

\State $MSE_{rd}\leftarrow MSE'_{rd}$

\Else

\State break

\EndIf 

\EndWhile 

\State \textbf{return} $MSE_{d}(a_{d,2})+MSE_{rd}$

\EndProcedure

\Procedure{OneIter}{$h_{k,r}$, $h_{k,d}$, $a'_{r,1}$, $a_{d,2}$}

\For {$k\in N_{r}$}\Comment{Iteration on $N_{r}$}

\State 

\State $\rho_{k}(P_{max})=\sqrt{P_{max}}/\sqrt{\left(a'_{r,1}h_{k,r}\right)^{2}+\left(a_{d,2}h_{k,d}\right)^{2}}$ 

\State $\gamma_{k}(P_{max})\negmedspace=\negmedspace\rho_{k}(P_{max})\cdot\left(\negmedspace\left(a'_{r,1}h_{k,r}\right)^{2}\negmedspace+\negmedspace\left(a_{d,2}h_{k,d}\right)^{2}\negmedspace\right)$

\If {$\gamma_{k}(P_{max})\ge1$ }

\State $\rho_{k}(P)=1/\left(\left(a'_{r,1}h_{k,r}\right)^{2}+\left(a_{d,2}h_{k,d}\right)^{2}\right)$

\State $b_{k,1}=\rho_{k}(P)\cdot a'_{r,1}h_{k,r}$

\State $b_{k,2}=\rho_{k}(P)\cdot a{}_{d,2}h_{k,d}$

\State $e_{k}=0$

\Else

\State $b_{k,1}=\rho_{k}(P_{max})\cdot a'_{r,1}h_{k,r}$

\State $b_{k,2}=\rho_{k}(P_{max})\cdot a{}_{d,2}h_{k,d}$

\State $e_{k}=1-\gamma_{k}(P_{max})$ 

\EndIf 

\EndFor

\State $MSE_{rd}=\sum_{k\in N_{r}}e_{k}^{2}+\sigma^{2}|a'_{r,1}|^{2}$

\State \textbf{return $MSE_{rd}$} 

\EndProcedure

\end{algorithmic}

\caption{\label{alg:relay2}Find optimal parameters and compute MSE.}
\end{algorithm}

In CohRelay, $a_{d,2}h_{k,d}b_{k,2}+a'_{r,1}h_{k,r}b_{k,1}\approx1$
so $a'_{r,1}h_{k,r}b_{k,1}$ is less than 1. This helps to reduce
the relay transmission power in Eq.\ref{eq:relay-pow}.

\subsection{Simulation Evaluation\label{sec:Evaluation}}

Here, we evaluate the relay methods discussed in the previous sections,
by comparing them with the AirComp method \cite{Liu20} that only
exploits the direct link. 

\begin{figure}
\centering

\includegraphics[width=9cm]{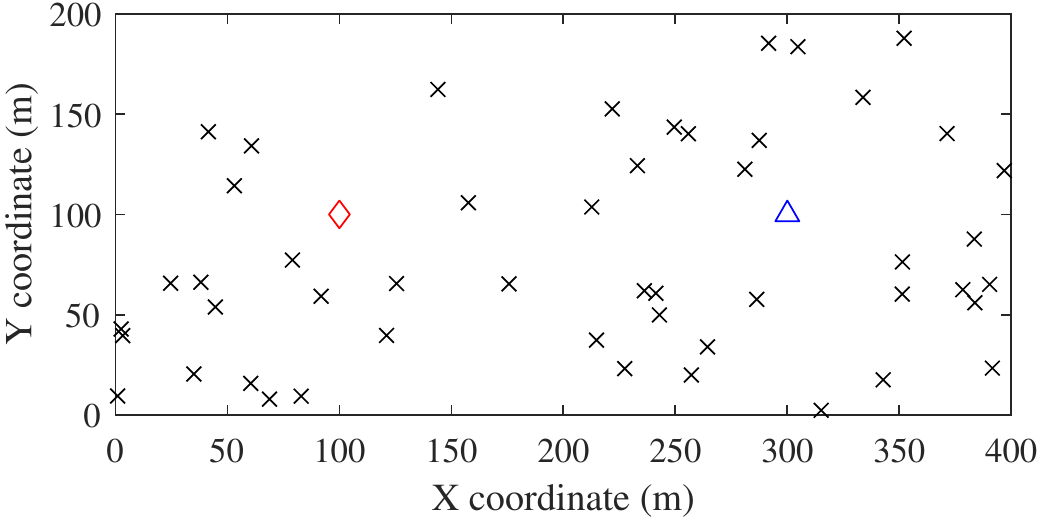}

\caption{\label{fig:scenario}An illustration of simulation scenario with 50
nodes ($\times$) randomly distributed in a 400m x 200m area, 1 sink
($\diamondsuit$, (100, 100)) and 1 relay ($\triangle$, (300, 100)).
Node deployment changes per evaluation.}
\end{figure}

Figure~\ref{fig:scenario} shows the simulation scenario. 50 sensor
nodes are randomly and uniformly distributed in a rectangle area.
The path loss model uses a hybrid free-space/two-ray model and each
link experiences independent slow Rayleigh fading (channel gains are
the same in two slots). We assume that the link $rd$ does not experience
fading by properly selecting a relay node not in fading (the relay
selection itself is left as future work). The noise level is -90dBm.
It is assumed that both sink $d$ and relay $r$ amplifies the signal
with a gain of 90dB. The simulation is run 10,000 times using the
Matlab software. Main parameters are listed in Table \ref{tab:SimCond}.

\begin{table}[!t]
\caption{Main parameters for evaluation.}

\label{tab:SimCond} 

\begin{centering}
\begin{tabular}{l|l}
\hline 
Term & Value\tabularnewline
\hline 
Area & 400m x 200m\tabularnewline
\# nodes & 50\tabularnewline
Frequency & 2.4GHz\tabularnewline
Channel & Free-space/two-ray, Rayleigh fading\tabularnewline
Power & $P_{max}=10$, $\sigma^{2}=1$\tabularnewline
\hline 
\end{tabular}
\par\end{centering}
\end{table}

\begin{figure}
\centering

\includegraphics[width=8cm]{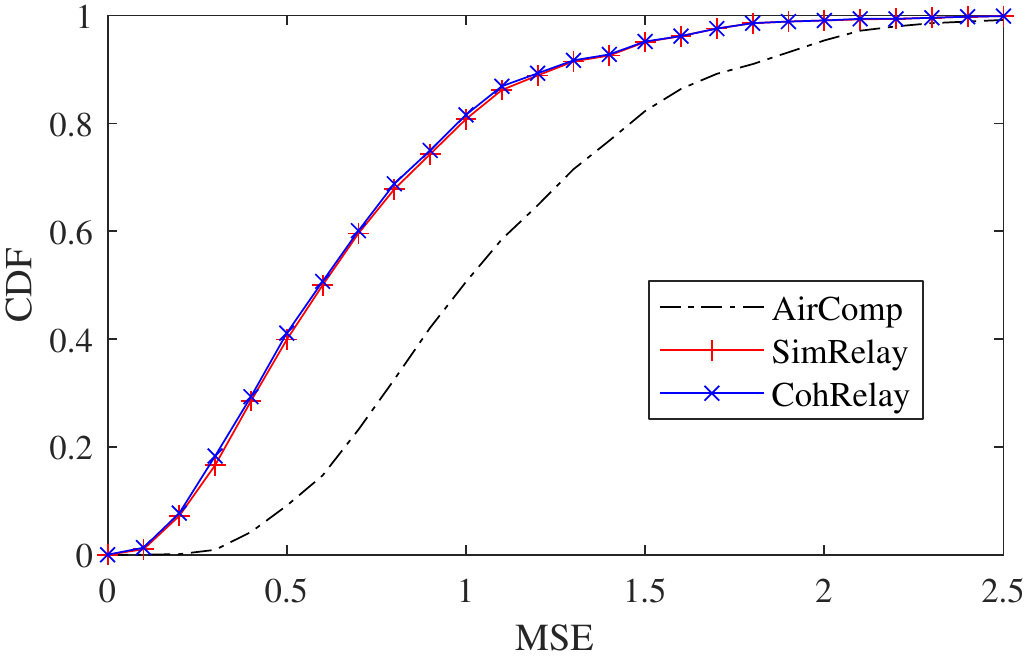}

(a) CDF of computation MSE

\includegraphics[width=8cm]{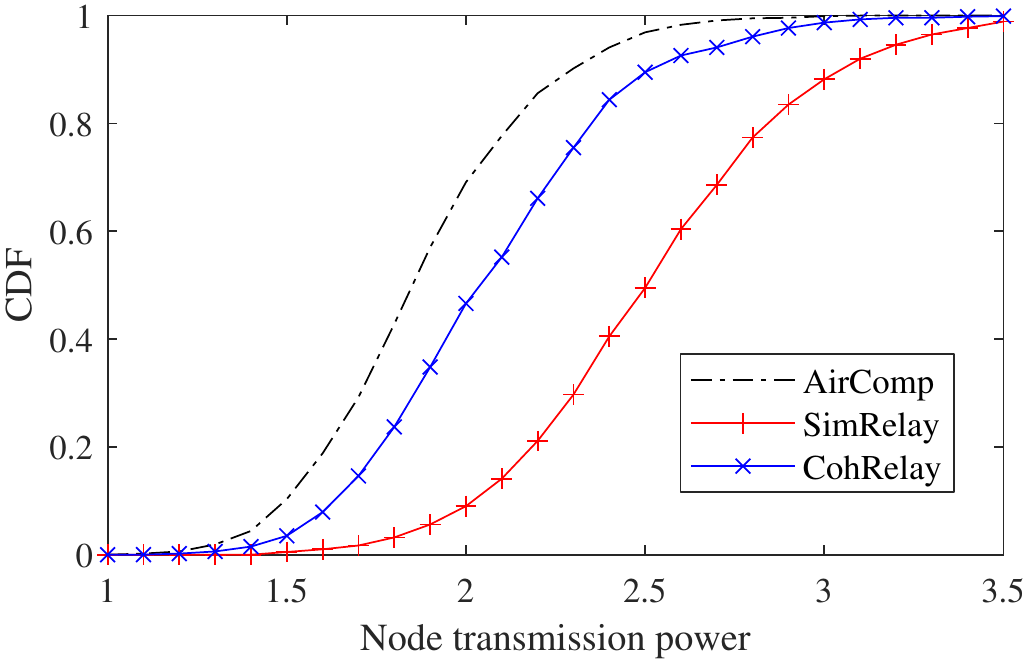}

(b) CDF of average node transmission power

\includegraphics[width=8cm]{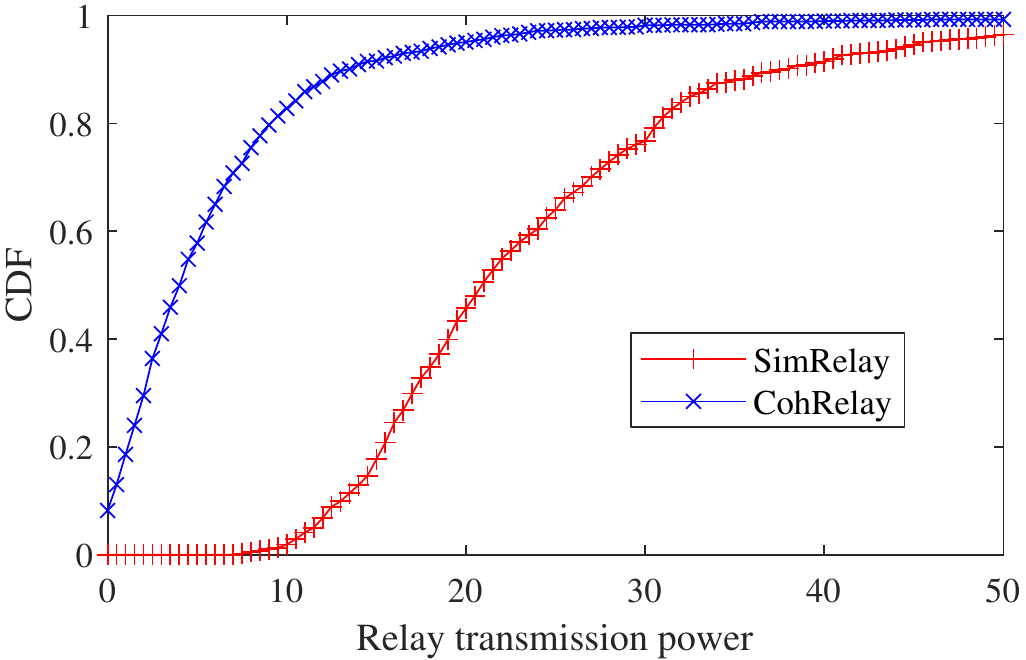}

(c) CDF of relay transmission power

\caption{\label{fig:rescdf}Cumulative distribution function of computation
MSE, average node transmission power, and relay transmission power
in different methods (30\% nodes use relay).}
\end{figure}

Figure~\ref{fig:rescdf} shows the cumulative distribution functions
(CDF) of computation MSE, average node transmission power and relay
transmission power in different methods. Obviously, AirComp using
only direct links has much larger computation MSE than relay methods.
SimRelay and CohRelay have almost the same performance in reducing
the computation MSE, but CohRelay has much smaller relay transmission
power than SimRelay. Surprisingly, both CohRelay and SimRelay require
more node transmission power than AirComp. 

\section{Refinement of the Relay Methods\label{sec:Refine}}

In the following, we analyze the computation MSE in CohRelay and refine
the parameters. A similar analysis applies to SimRelay. 

In Eq.(\ref{eq:mse-relay2}), the computation MSE is composed of the
signal part and the noise part. When reducing the computation MSE,
at first mainly MSE of the signal part is reduced to align signal
magnitudes. When MSE of the signal part gets small and the noise becomes
dominant, MSE of the noise part is also reduced, which leads to smaller
$a'_{r,1}$ and $a_{d,2}$, and increased power ($b_{k,1}$ and $b_{k,2}$). 

In the original AirComp, MSE of the noise part is $a^{2}\sigma^{2}$.
To avoid over-reducing MSE of the noise part in CohRelay, we restrict
$((a'_{r,1})^{2}+d_{d,2}^{2})\sigma^{2}$ to be no less than $\gamma\cdot a^{2}\sigma^{2}$,
where $\gamma$ is a parameter (later it is set to 1.0 based on simulation
results). In addition, MSE of the signal part is not a continuous
function. Actually it is a constant when $a'_{r,1}$ and $a_{d,2}$
change within a range, because the variation is absorbed by adjusting
$b_{k,1}$ and $b_{k,2}$, which change continuously. Then, a small
decrease in the computation MSE may lead to a large increase in transmission
power. To capture this feature, we consider a new metric, involving
both the computation MSE and node transmission power (TxP), as follows,
where $\theta$ is a parameter.

\begin{align}
\underset{a'_{r,1},a_{d,2},(a'_{r,1})^{2}+a_{d,2}^{2}\ge\gamma\cdot a^{2}}{argmin}\theta\cdot MSE+(1-\theta)\cdot TxP,\label{eq:metric}\\
TxP=\frac{1}{K}\sum_{k\in N_{r}\cup N_{d}}b_{k,1}^{2}+b_{k,2}^{2}.\nonumber 
\end{align}
With each candidate $a_{d,2}$ in a certain range, $a'_{r,1}$ is
computed from $\sqrt{\gamma\cdot a^{2}-a_{d,2}^{2}}$. With $a'_{r,1}$
and $a_{d,2}$, transmission power ($b_{k,1}$, $b_{k,2}$) for node
$k\in N_{r}$ is computed using the function OneIter in Algorithm
\ref{alg:relay2}. For node $k\in N_{d}$, $b_{k,2}$ is computed
from $a_{d,2}$. Then, the computation MSE and average TxP are computed.
Because their values are of the same order of magnitude, $\theta$
is set to 0.5.

\begin{figure}
\centering

\includegraphics[width=8cm]{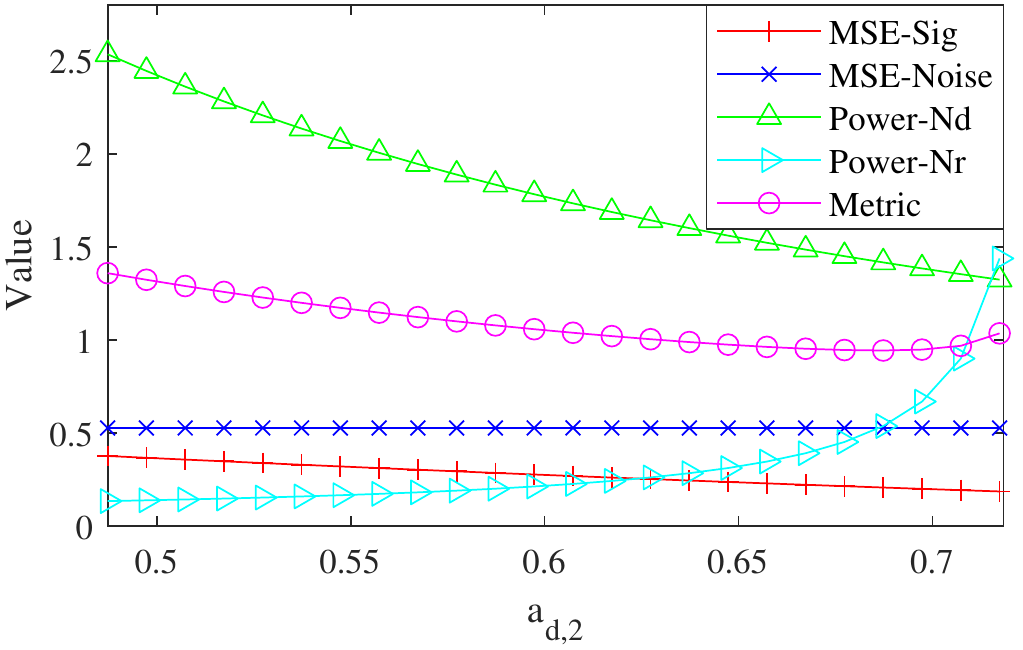}

\caption{\label{fig:metric}Variation of different metrics with $a_{d,2}$
($\gamma=1.0$).}
\end{figure}

With this new metric, we evaluate MSEs of the signal part and the
noise part, average transmission power of nodes in $N_{r}$ and $N_{d}$,
and the metric in Eq.(\ref{eq:metric}). The results are shown in
Fig.\ref{fig:metric}. MSE of noise part is kept constant, as expected.
With the increase of $a_{d,2}$, transmission power of nodes in $N_{d}$
decreases. Meanwhile $a'_{r,1}$ decreases, which leads to a quick
increase in transmission power of nodes in $N_{r}$ when $a_{d,2}$
is large ($a'_{r,1}$ is small). The overall MSE of the signal part
decreases. Then, the metric, as a weighted sum of the overall MSE
and average power, reaches a minimum somewhere, which prefers to use
a smaller transmission power when the computation MSE has no significant
change.

\begin{figure}
\centering

\includegraphics[width=8cm]{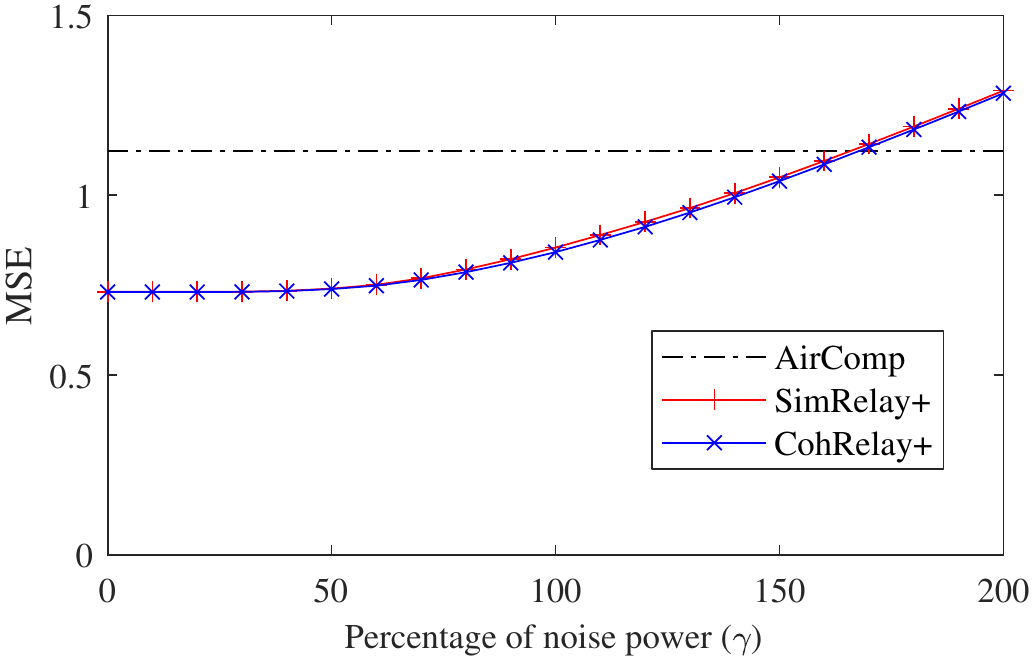}

(a) Computation MSE

\includegraphics[width=8cm]{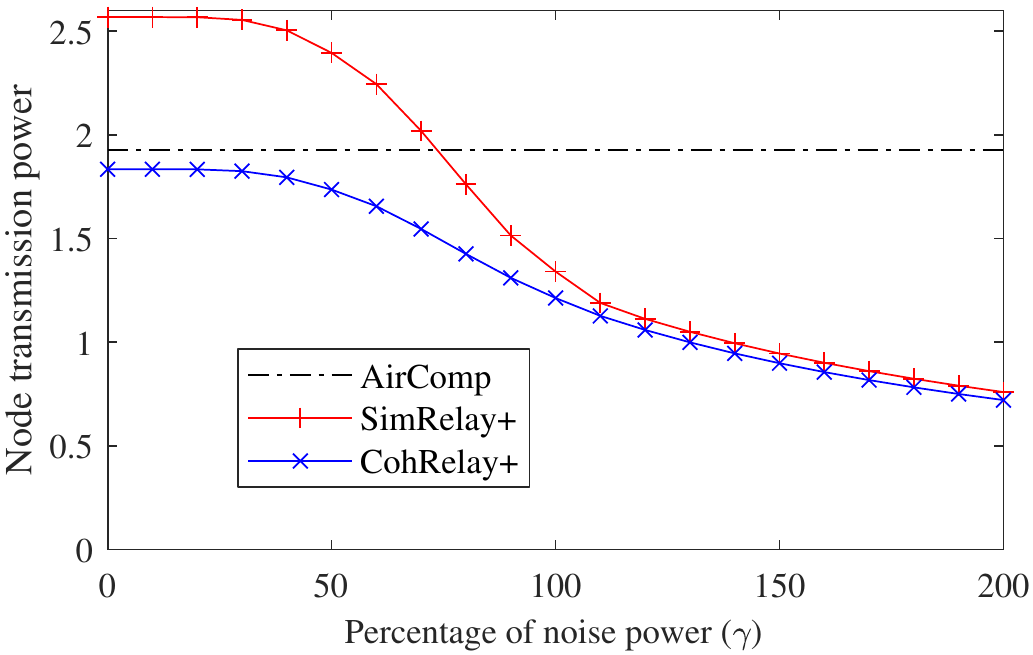}

(b) Average node transmission power

\includegraphics[width=8cm]{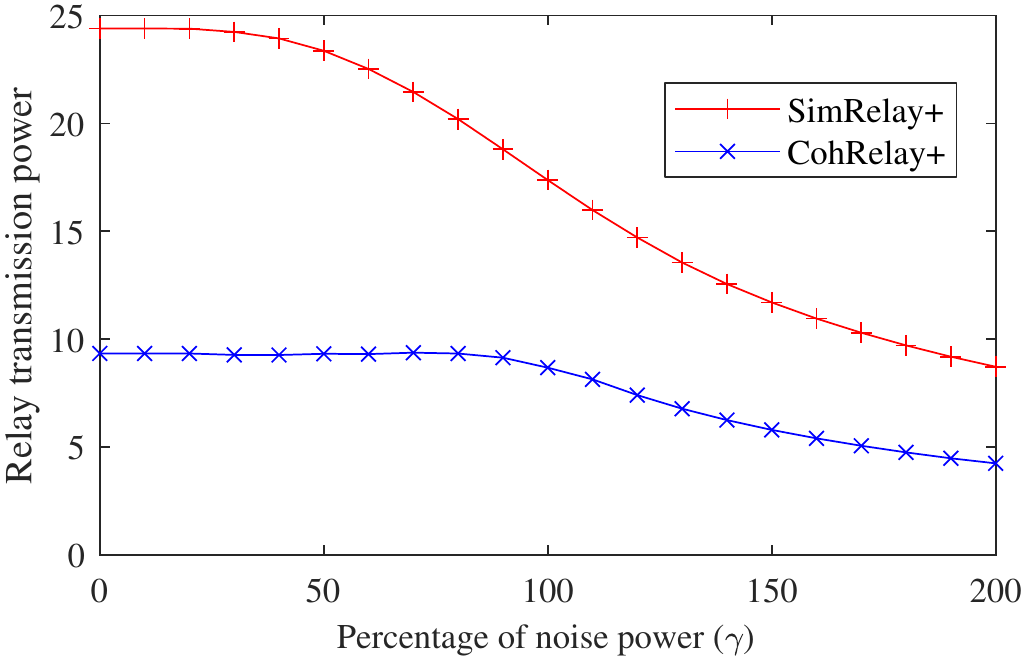}

(c) Relay transmission power

\caption{\label{fig:noise-ratio}Computation MSE, average node transmission
power, and relay transmission power in different methods. (30\% nodes
use relay).}
\end{figure}

Next, we evaluate the computation MSE, node transmission power, and
relay transmission power, by changing the parameter $\gamma$. Hereafter,
the refined relay methods are renamed as SimRelay+ and CohRelay+,
respectively. The results are shown in Fig.\ref{fig:noise-ratio}.
At $\gamma=200\%$, it is equivalent that noise at the relay and the
sink are directly added together. When $\gamma$ decreases from 200\%
to 100\%, MSE of the noise part is also reduced, so the overall MSE
gradually decreases, meanwhile node and relay transmission power increases,
although slowly. When $\gamma$ further decreases, the decrease in
the computation MSE becomes smaller while the increase in node transmission
power becomes larger. Therefore, in the following evaluation, $\gamma$
is set to 100\%. 

\begin{figure}
\centering

\includegraphics[width=8cm]{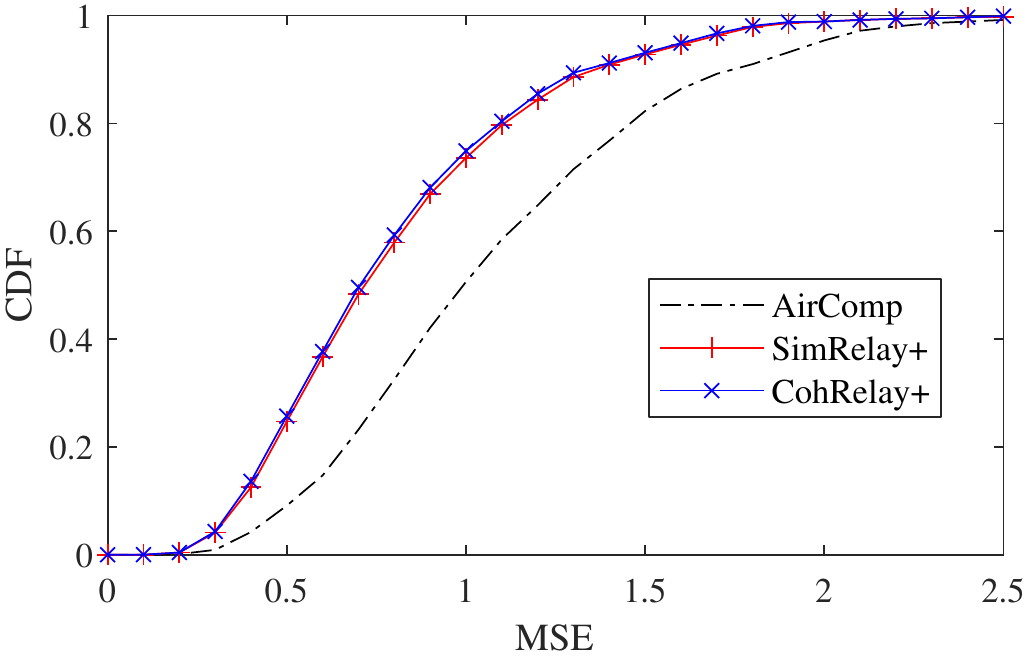}

(a) CDF of computation MSE

\includegraphics[width=8cm]{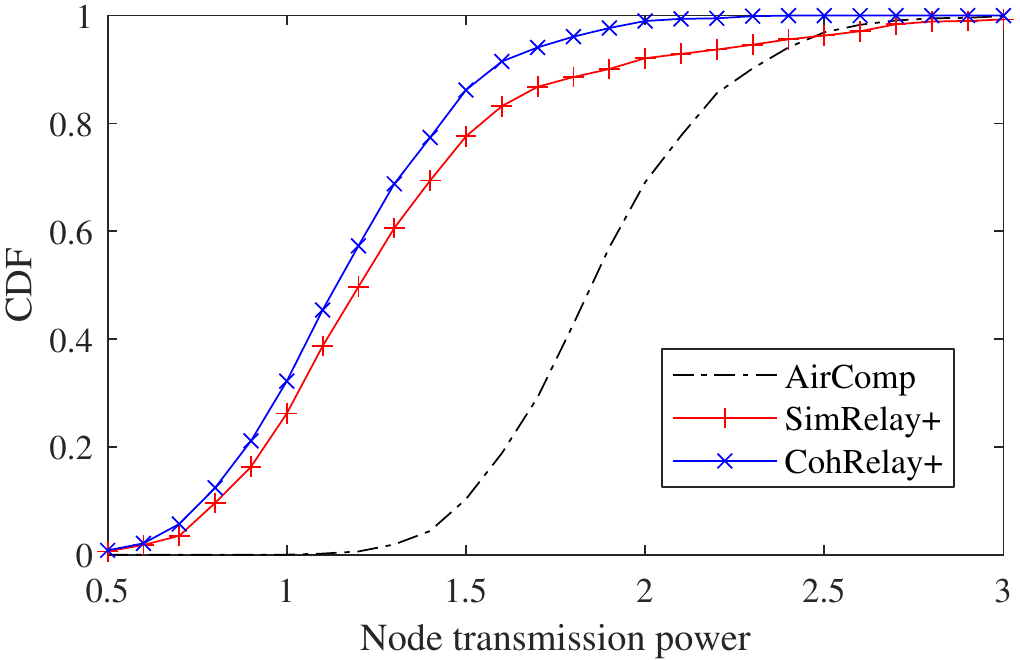}

(b) CDF of average node transmission power

\includegraphics[width=8cm]{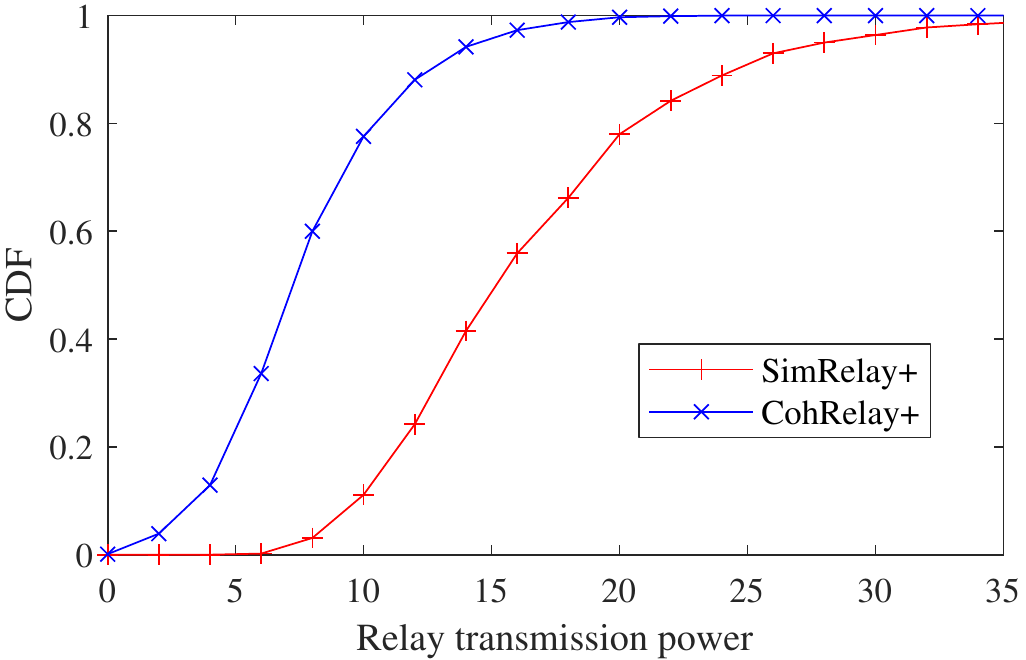}

(c) CDF of relay transmission power

\caption{\label{fig:rescdf-new}Cumulative distribution function of computation
MSE, average node transmission power, and relay transmission power
in different methods. (30\% nodes use relay. $\gamma=100\%$).}
\end{figure}

With $\gamma=100\%$, we re-evaluate the computation MSE, node and
relay transmission power. The results are shown in Fig.\ref{fig:rescdf-new}.
The reduction of the computation MSE in CohRelay+ compared with AirComp,
decreases from 35.6\% (Fig.\ref{fig:rescdf}(a)) to 25.0\%. But the
reduction of average node transmission power in CohRelay+ is greatly
improved from -10.0\% (Fig.\ref{fig:rescdf}(b)) to 37.0\%.

\begin{figure}
\centering

\includegraphics[width=8cm]{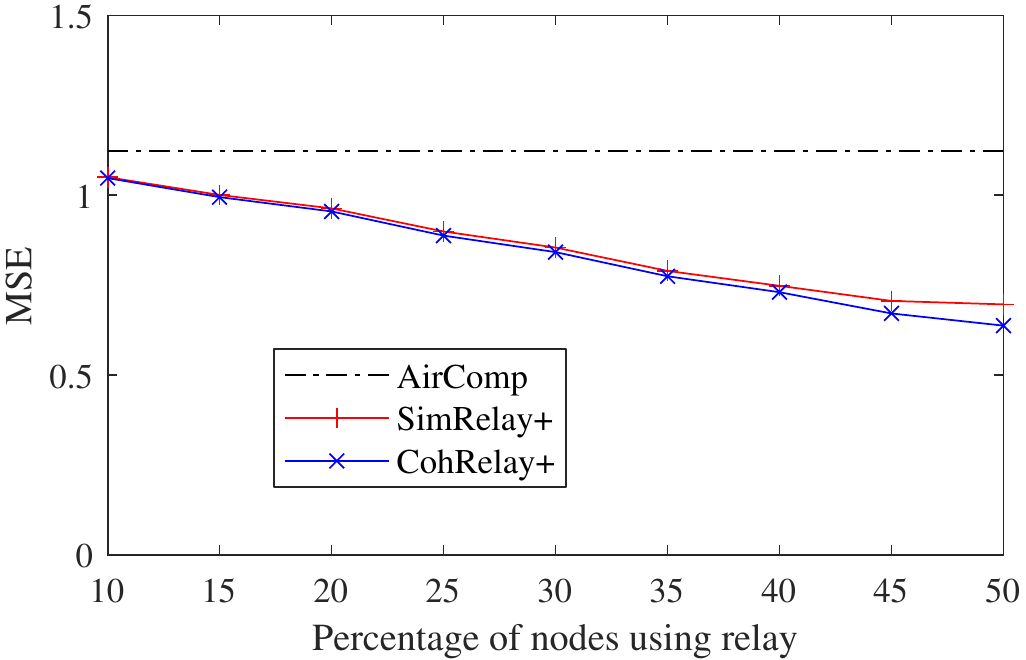}

(a) Computation MSE

\includegraphics[width=8cm]{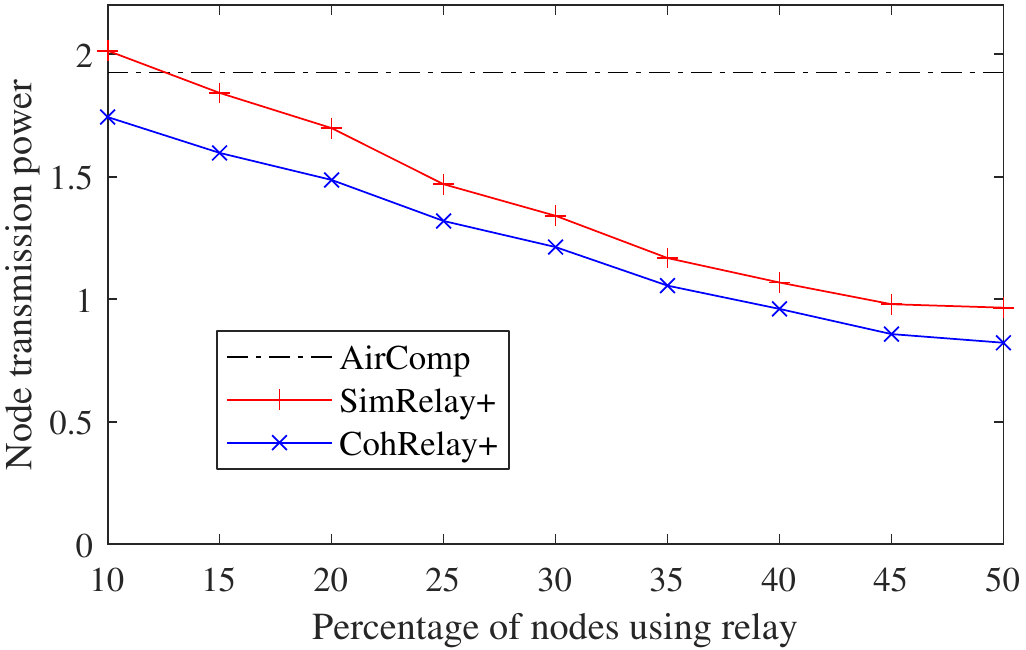}

(b) Average node transmission power

\includegraphics[width=8cm]{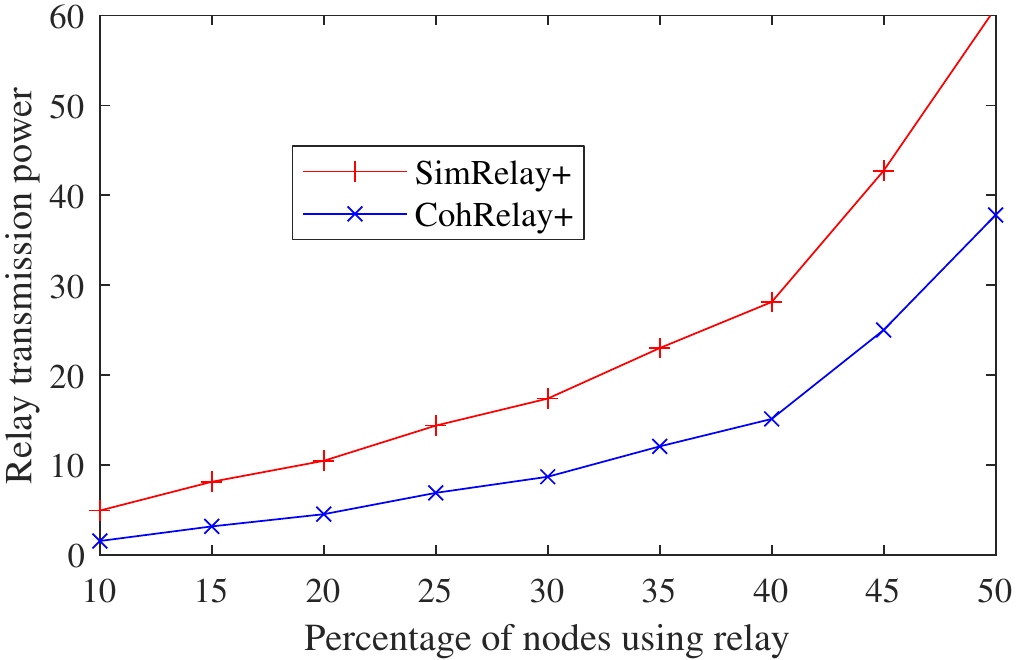}

(c) Relay transmission power

\caption{\label{fig:relay-ratio}Computation MSE, average node transmission
power, and relay transmission power with respect to different percentages
of nodes using the relay ($\gamma=100\%$).}
\end{figure}

We further investigate the impact of the percentage of nodes using
relay. As shown in Fig.\ref{fig:relay-ratio}, when only a small percentage
of nodes use the relay, the reduction of the computation MSE is limited.
Node transmission power is relatively large but the relay transmission
power is small. When more nodes use relay, the computation MSE is
further reduced, so is node transmission power, but relay transmission
power increases. In all cases, CohRelay+ achieves almost the same
(or a little smaller) computation MSE as SimRelay+, but reduces node
transmission power, and especially reduces the relay transmission
power by half or more when the percentage of nodes using relay is
less than 30\%. In this range, the relay transmission power in CohRelay+
is less than $P_{max}$, which makes it practical to use a relay node.

\begin{figure}
\centering

\includegraphics[width=8cm]{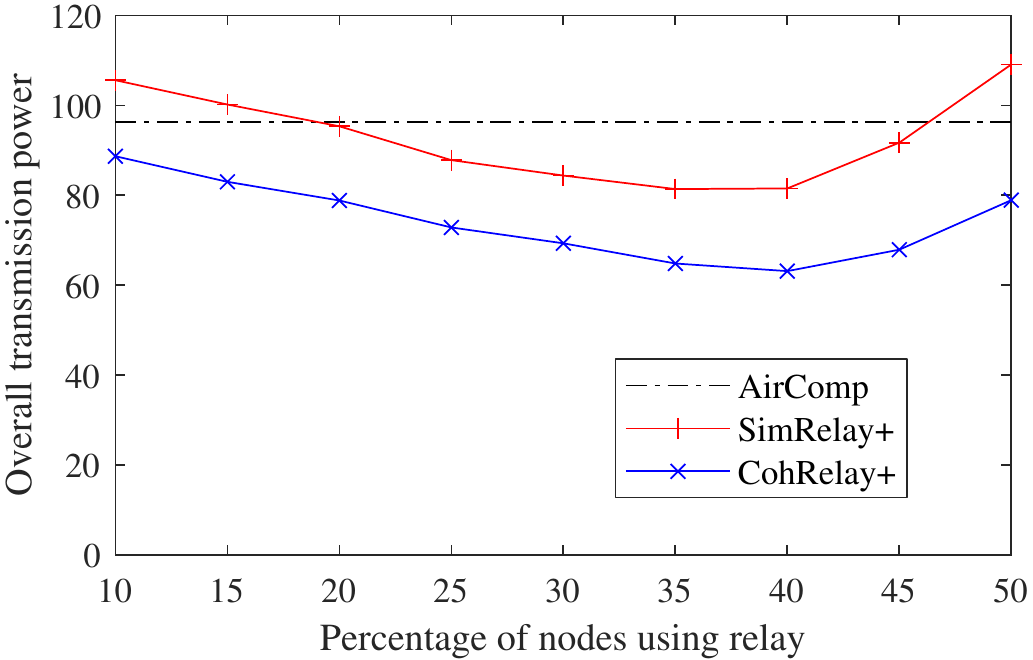}

\caption{\label{fig:powall-ratio}Overall transmission power of all nodes and
the relay, under different percentages of nodes using the relay ($\gamma=100\%$).}
\end{figure}

In the above evaluation, node transmission power and relay transmission
power are separately evaluated, and a relay is not used in AirComp.
For a fair comparison, we further investigate the overall transmission
power of all nodes and the relay. The result is shown in Fig.\ref{fig:powall-ratio}.
With the increase of the percentage of nodes using relay, the overall
transmission power in SimRelay+ and CohRelay+ decreases at first,
because using relay helps to reduce node transmission power, and then
increases because of the large transmission power at the relay. When
the percentage of nodes using relay is no more than 50\%, CohRelay+
consumes less overall transmission power than AirComp. 

In sum, using a relay, SimRelay+ and CohRelay+ reduce node transmission
power and have a similar performance in reducing the computation MSE,
compared with AirComp. This is achieved at the cost of one more slot,
and potentially more transmission power. As for the overall transmission
power, SimRelay+ may consume more power, but CohRelay+ always consume
less power than AirComp in the typical range (the percentage of nodes
using relay is less than 50\%). Compared with SimRelay+, CohRelay+
reduces the relay transmission power by half or even more, to below
the limit when the percentage of nodes using relay is no more than
30\%, which facilitates the practical application of AirComp.

\section{Conclusion \label{sec:Conclusion}}

AirComp greatly improves the efficiency of data collection and processing
in sensor networks. But its performance is degraded when signals of
nodes far away from the sink cannot arrive at the sink, aligned in
signal magnitude. To address this problem, this paper investigates
the amplify and forward based relay method, and discusses practical
issues such as the large relay transmission power and the over-increase
of node transmission power. As for the two relay polices, SimRelay+,
being simple, effectively reduces the computation MSE and node transmission
power. With coherent combination of direct signals and relayed signals,
CohRelay+ further reduces the node transmission power and relay transmission
power. Although it is impractical for all nodes to use the relay,
CohRelay+ helps to reduce the computation MSE meanwhile keeping the
relay transmission power below the limit by adjusting the percentage
of nodes using relay.  In the future, we will further study the relay
selection problem.

\bibliographystyle{IEEEtran}
\bibliography{IEEEabrv,mybibfile}

\end{document}